\documentclass[prb,twocolumn,showpacs,preprintnumbers,amsmath,amssymb]{revtex4}

\usepackage{graphicx}
\usepackage{dcolumn}
\usepackage{bm}
\usepackage{mathrsfs}
\usepackage{float}
\usepackage{wrapfig}
\usepackage{color}

\begin{document}


\title{Doublon Dynamics of the Hubbard Model on Triangular Lattice}

\author{Toshihiro Sato$^1$}
\author{Hirokazu Tsunetsugu$^2$}
\affiliation{
$^1$Computational Condensed Matter Physics Laboratory, RIKEN, 2-1 Hirosawa, Wako, Saitama 351-0198, Japan\\
$^2$Institute for Solid State Physics, University of Tokyo, 5-1-5 Kashiwanoha, Kashiwa, Chiba 277-8581, Japan}

\date{\today}

\begin{abstract}

We study  the dynamics of doublon in the half-filled Hubbard model on the triangular lattice
by using the cellular dynamical mean field theory. 
Investigating the nearest-neighbor dynamical correlations, we demonstrate that a nearest-neighbor doublon-holon pair shows a strong attraction,
in particular in the insulating phase.
We also calculate the on-site dynamical correlation of doublon and find that the life time of doublon is longer in the metallic phase
than in the insulating phase. 
In the long-time region, the metallic phase has persistent fluctuations in various nearest-neighbor configurations,
while the fluctuations are vanishingly small in the insulating phase.
Obtained results indicate clear differences of dynamics of doublon between in the metallic and in the insulating phases.
\end{abstract}

\maketitle


Since the observation of the Mott metal-insulator transition in Cr-doped V$_2$O$_3$,\cite{MT-exp-1}
many properties of the Mott transition have been studied from theoretical and experimental aspects.
Recent hot topics of the Mott transition include critical properties\cite{MT-critexp-1,MT-critexp-2,MI-crit-1,MTC-dcc-DMFT-1,docc-scaling,docc-scaling-2}
and dynamical characteristics.~\cite{rf-1,rf-2,rf-3,rf-4,rf-5,rf-6,rf-7}
It is well known that doublon (doubly occupied site) plays the role of order parameter in the Mott transition.\cite{docc-op}
Some theoretical groups have investigated behaviors of doublons upon changing the electron correlation near the Mott transition.
Understanding of its thermodynamic criticality has also made progress by performing a scaling analysis of 
a singularity in doublon density near the critical end point of the Mott transition.\cite{MI-crit-1,MTC-dcc-DMFT-1,docc-scaling,docc-scaling-2}
Correlations between doulon and holon (vacant site) have been also studied theoretically and it was proposed that they form a bound state
in the insulating phase and the Mott transition is characterized by its binding and unbinding.\cite{dc-3,dc-4}
To study dynamical properties in the Mott transition, dynamical spin or charge susceptibility~\cite{rf-1,rf-2,rf-7,rf-3} and 
optical conductivity~\cite{docc-scaling-2,rf-4,rf-5,rf-6} have been calculated for the Hubbard models by using
the dynamical mean field theory (DMFT).~\cite{DMFT}
A clear difference has been reported for the spin and charge dynamics
between in the metallic and in the insulating phases.
However, the dynamics of doublons is not well understood.
This is the main issue of this paper and we will report our numerical study on the dynamical properties of doublons in the triangular-lattice Hubbard model.

Original Mott transition occurs inside the paramagnetic phase at a finite temperature without magnetic transition
and its realization involves magnetic frustration.
In this letter, we numerically study the dynamics of doublons and holons in a frustrated Hubbard model.
Using cellular dynamical mean field theory (CDMFT),~\cite{CDMFT} we calculate and examine their dynamical correlations.

%
To this end, we employ the half-filled triangular-lattice Hubbard model.
Its Hamiltonian reads as
\begin{eqnarray}
H=-v\sum_{\langle i,j \rangle,\sigma}c_{i\sigma}^\dagger c_{j\sigma}+U\sum_{i}n_{i\uparrow}n_{i\downarrow}-\mu\sum_{i,\sigma}n_{i\sigma}.
\label{eq:H-1}
\end{eqnarray}
In this model, the transition occurs inside the paramagnetic region.~\cite{magphase-trian}  
Note that the nearest-neighbor hopping integral is denoted by $v$ and the symbol $t$  is reserved for time.
 $U$ is the on-site Coulomb repulsion and the chemical potential $\mu$ tunes electron density to half filling.
$ c_{i\sigma}^\dagger(c_{i\sigma})$ is an electron creation (annihilation) operator at site $i$ with spin $\sigma$
and electron density operator is $n_{i\sigma}=c_{i\sigma}^\dagger c_{i\sigma}$.
We use the CDMFT with a three-site triangular cluster to calculate dynamical correlations.
We numerically obtain the single- and two-electron Green's functions inside the cluster by using the continuous-time quantum Monte Carlo (CTQMC) solver based on the strong coupling expansion.\cite{CTQMC}

Dynamical correlations are defined for real time $t$ by,
\begin{eqnarray}
S^{j}_{\rm oo'}(t)&=&\langle \hat{o}_{1}(t)\hat{o}'_{j}(0)\rangle,
\label{eq:dsf}
\end{eqnarray}
where the average is calculated for thermal equilibrium at temperature $T$.
$\hat{o}_{j}(\hat{o}'_{j}) $ are operators at site $j$ inside the cluster of doublon density $\hat{d}_{j}$ or holon density  $\hat{h}_{j}$
or single-occupied density $\hat{s}_{j}$ described as,
\begin{eqnarray}
\hat{d}_{j}=n_{j\uparrow}n_{j\downarrow}, \hat{h}_{j}=(1-n_{j\uparrow})(1-n_{j\downarrow}),\nonumber \\
\hat{s}_{j}=1-(\hat{d}_{j}+\hat{h}_{j})=n_{j\uparrow}+n_{j\downarrow}-2\hat{d}_{j}.
\label{eq:de}
\end{eqnarray}
In our results, we have checked that the three sites inside the cluster are always all equivalent, and therefore it does not loose generality to choose the site 1
for one quantity $\hat{o} $.
For the other quantity $\hat{o}'$, it is relevant only if the site $j$ is 1 or not.
The $j$$=$$1$ case corresponds to on-site correlations and the $j$$=$$2$ case corresponds to nearest-neighbor correlations, while the choice of $j$$=$$3$ is completely equivalent to $j$$=$$2$.
Note also that the dynamic correlations $S^{j}_{\rm oo'}(t)$ have generally complex values.
Therefore, for their interpretation, it is more convenient to see,
\begin{eqnarray}
\Gamma^{j}_{\rm oo'}(t)&=&|S^{j}_{\rm oo'}(t)|-|\langle \hat{o}_{1}\rangle\langle \hat{o}'_{j}\rangle|,
\label{eq:dsf2}
\end{eqnarray}
rather than more conventional $|S^{j}_{\rm oo'}(t)-\langle \hat{o}_{1}\rangle\langle \hat{o}'_{j}\rangle|$, since it is real and its sign has meaning:
$\Gamma^{j}_{\rm oo'}(t)$$>$$0$ means attractive correlation, and  $\Gamma^{j}_{\rm oo'}(t)$$<$$0$ means repulsive correlation. 

The following is a brief sketch of numerical procedures.
In the CDMFT, we obtain the corresponding correlation function for Matsubara frequency $S^{j}_{\rm oo'}(i\omega_n)$
by averaging over 512 imaginary-time MC samples.
We then perform analytic continuation $i\omega_{n}$$\rightarrow$$\omega$$+$$i 0$ based on the maximum entropy algorithm (MEM)\cite{MEM}
to calculate real frequency quantity $S^{j}_{\rm oo'}(\omega)$.
We finally perform Fourier transformation to obtain $S^{j}_{\rm oo'}(t)$.
In what follows, we normalize $\omega$, $t$, $U$, and $T$ by the energy unit $v$.

\begin{figure}
\centering
\vspace{-3.5cm}
\centerline{\includegraphics[height=5.25in]{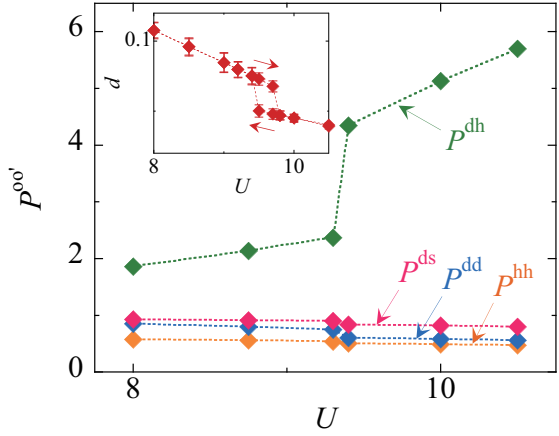}}
\vspace{-3cm}
\caption{(Color online) 
$U$-dependence of nearest-neighbor equal-time correlations defined by eq.~(\ref{eq:Poo}) at $T$=$0.08$.
(Inset) $U$-dependence of doublon density at $T$=$0.08$.
}  
\label{fig:3}
\end{figure}

\begin{figure}
\centering
\vspace{0cm}
\vspace{-3.5cm}
\centerline{\includegraphics[height=5in]{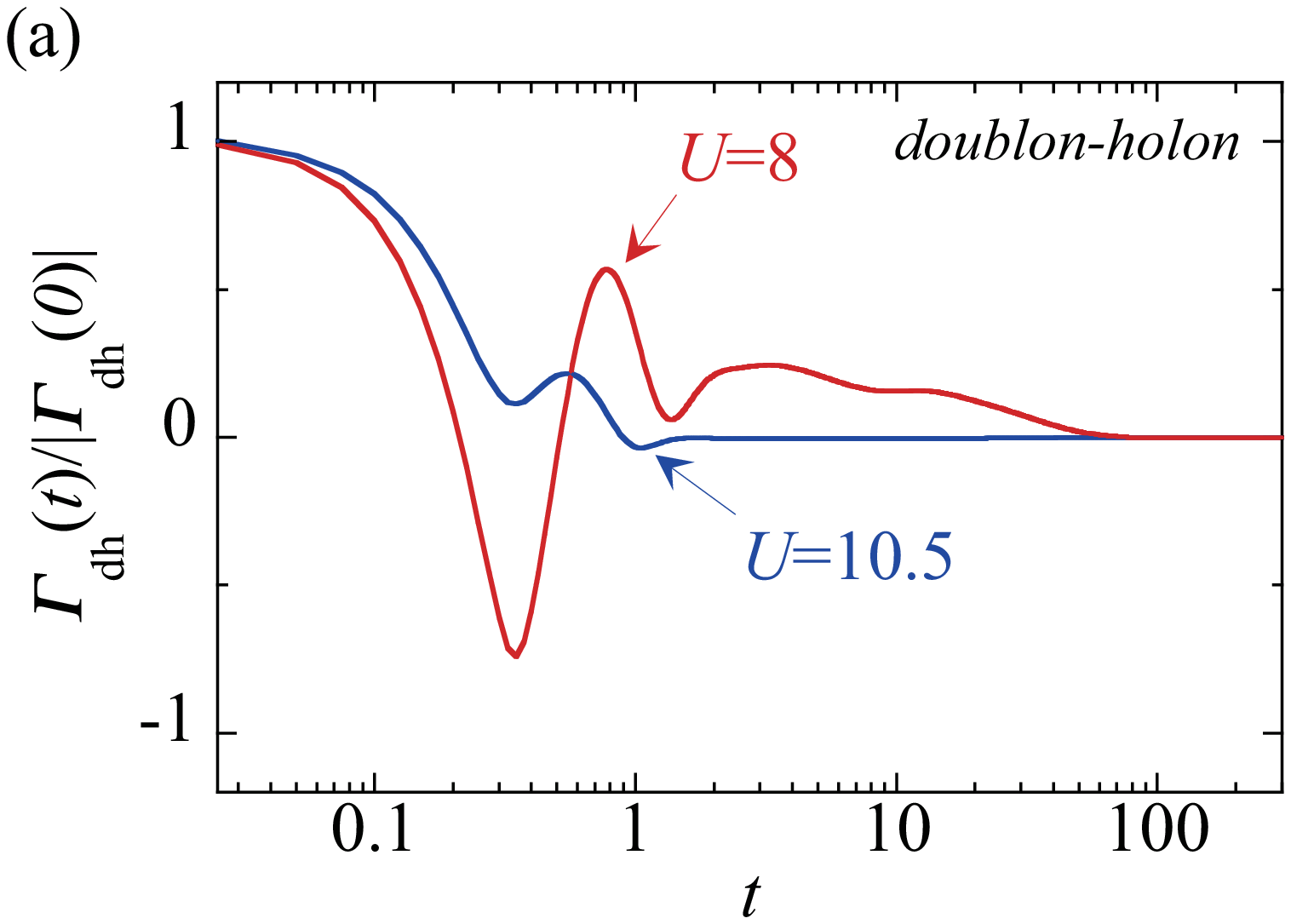}}
\vspace{-7cm}
\centerline{\includegraphics[height=5in]{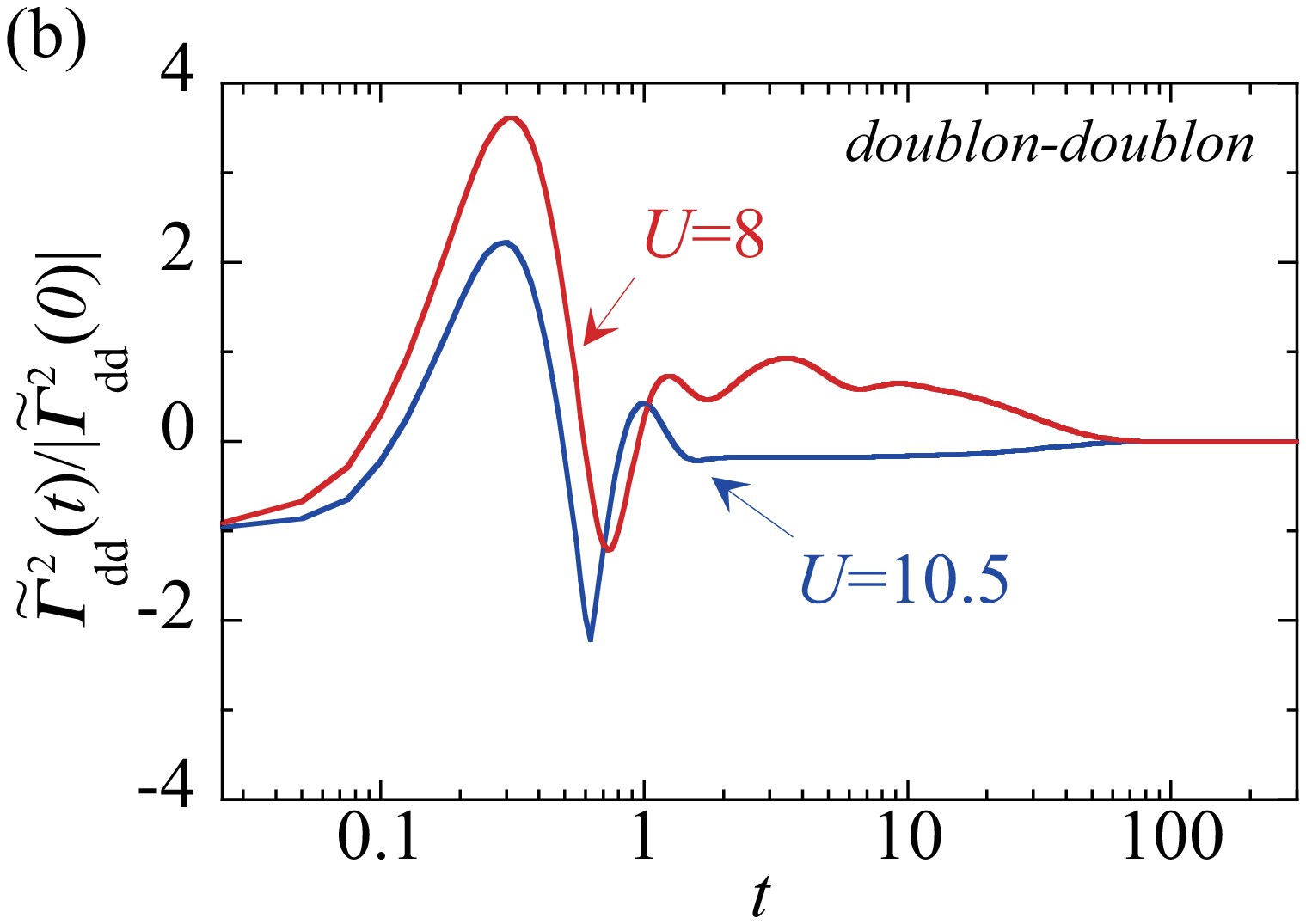}}
\vspace{-6.75cm}
\centerline{\includegraphics[height=5in]{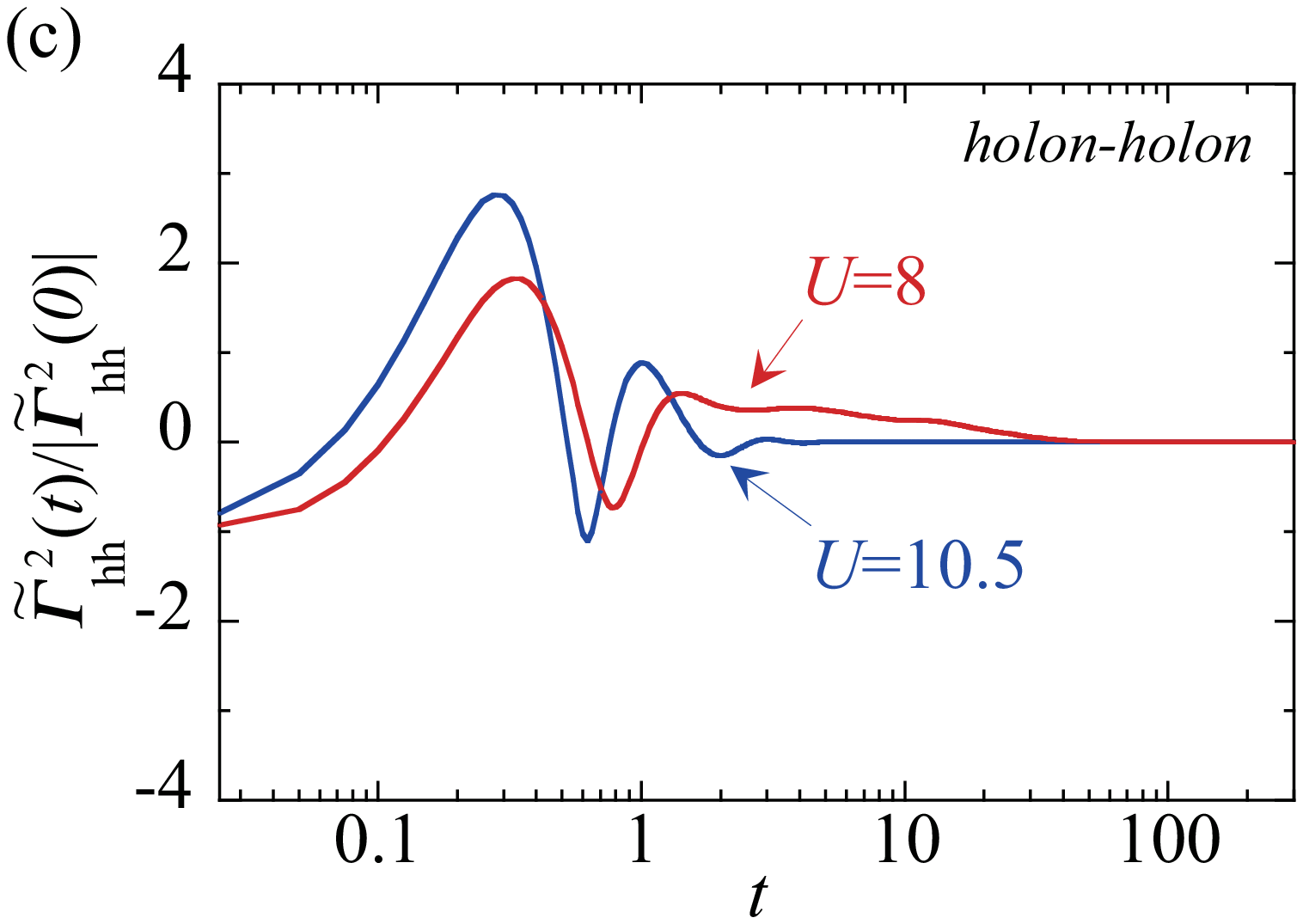}}
\vspace{-3.25cm}
\caption{(Color online) 
Nearest-neighbor dynamical correlations at $T$=$0.08$ between (a) doublon and holon,
(b) two doublons, and (c) two holons.
}  
\label{fig:4}
\end{figure}

%
We start with confirming the finite-temperature Mott transition in the parameter space of $U$ and $T$.
We investigate the $U$-dependence of the doublon density, $d\equiv\frac{1}{N_{\rm c}}\sum_{i}\langle \hat{d}_{i}\rangle$, for various $T$'s.
Here, $N_{\rm c}=3$ is the cluster size.
Note that the half-filling condition determines the densities of holon and singly occupied sites as $h$$=$$d$, $s$$=$$1$$-$$2d$.  
The inset of Fig.~\ref{fig:3} presents $d(U)$ at $T$=$0.08$.
$d(U)$ shows a jump and hysteresis, which is a characteristic of the first-order Mott transition.
At higher $T$=$0.15$, $d(U)$ shows no hysteresis and only a smooth change from insulator to metal. 
Analyzing the singularity in $d(U)$ at various $T$'s,  we determine the $U$$-$$T$ phase diagram
and this is consistent with the previous study.~\cite{phase} 
A line of the first-order Mott transition terminates at the critical end point, $U^* $$\sim$$ 9.4$ and $T^*$$ \sim$$ 0.10$.
In the following, we will investigate dynamics with decreasing $U$ at  the fixed temperature $T$=$0.08$ $<$ $T^*$,
where the first-order Mott transition occurs at $U_{\rm c}$$=$$9.4$.

We first examine the correlations between nearest-neighbor sites, $\Gamma^{2}_{\rm oo'}(t)$.
One important characteristics is the equal-time values, $t$$=$$0$.
We define their normalized correlation by, 
\begin{eqnarray}
P^{\rm oo'}=\frac{\langle \hat{o}_{1}\hat{o}'_{2}\rangle}{\langle \hat{o}_{1}\rangle\langle \hat{o}'_{2}\rangle},
\label{eq:Poo}
\end{eqnarray}
where $\hat{o}$ and $\hat{o'}$ are $\hat{d}$, $\hat{h}$, or $\hat{s}$.
$P^{\rm oo'}$$>$$1$ means an attractive equal-time correlation between nearest-neighbor sites,
while $P^{\rm oo'}$$<$$1$ means repulsive equal-time correlation.
The main panel of Fig.~\ref{fig:3} shows their $U$-dependence for typical combinations of configuration.  
The most important feature is a large value of $P^{\rm dh}$, and this manifests a strong nearest-neighbor attraction between doublon and holon.
This attraction is noticeably enhanced in the insulating phase.
An opposite behavior is found in the correlation between two doublons, between two holons, and between doublon and singly occupied site:
they are repulsive to each other.
Their repulsions are also enhanced in the insulating phase, but these enhancements are not as large
as the enhancement of doublon-holon attraction.
The difference between $P^{\rm hh}$ and $P^{\rm dd}$ is due to the particle-hole asymmetry of the model, since the lattice is not bipartite.
It is natural that $P^{\rm dd}$$>$$P^{\rm hh}$, because the low-energy density of states is larger for positive energy; i.e, adding electron is easier.  
 
Dynamical correlations also exhibit clear differences between in the metallic and in the insulating phases. 
Figures~\ref{fig:4} (a), (b), and (c) show the dynamics of doublon-holon pair, doublon pair, and holon pair, respectively,
in the metallic ($U$=$8$) and insulating ($U$=$10.5$) phases at $|U-U_c|$$\sim$$1$.
The characteristic time scale in the short time part is about $t$$\sim$$0.3$ in all cases, and this time scale is longer particularly for doublon-holon pair.
The short-$t$ behavior of this doublon-holon pair is opposite to the other two cases.
The doublon-holon pair decays into other configurations as $t$ increases, and the corresponding pair life time becomes longer in the insulating phase.
In contrast, the prohibited doublon pair and holon pair starts to be dynamically formed with retardation.
The formation of doublon pair occurs faster in the metallic phase than in the insulating phase,
while the formation of holon pair shows an opposite behavior.

Another important difference is long-$t$ behavior.
In the insulating phase, the correlations for $t$$>$$2$ is very small, almost vanishing.
The behavior in the metallic phase is in sharp contrast, and the correlations persist up to long time $t$$\sim$$50$$-$$60$. 
Moreover, we find that the $t$-dependence at smaller $U$$=$$8$ shows many structures.
These results may be due to the larger charge fluctuation in the metallic phase.

To examine these structures, we have also investigated the complex $S^{2}_{\rm oo'}(t)$ and have found that the structure at $t$$\sim$$0.3$
corresponds to its position with phase $\pi$ or $0$. 
To see more details, we have checked the correlations in the frequency domain ${S}^{2}_{\rm oo'}(\omega)$
as we will analyze later for the on-site doublon dynamics.
There are some low-$\omega$ peaks, in particular in the metallic phase, in addition to a broad peak around $\omega$=$U$.
The broad peak around $\omega$=$U$ corresponds to the incoherent dynamics by the excitations to the Hubbard band and
the short-$t$ dynamics is predominated by the contribution of the broad peak.
On the other hand, we have found that the low-$\omega$ peaks dominate the long-$t$ dynamics  and are absent in the insulating phase.

\begin{figure}
\centering
\vspace{-3cm}
\centerline{ \includegraphics[height=4.5in]{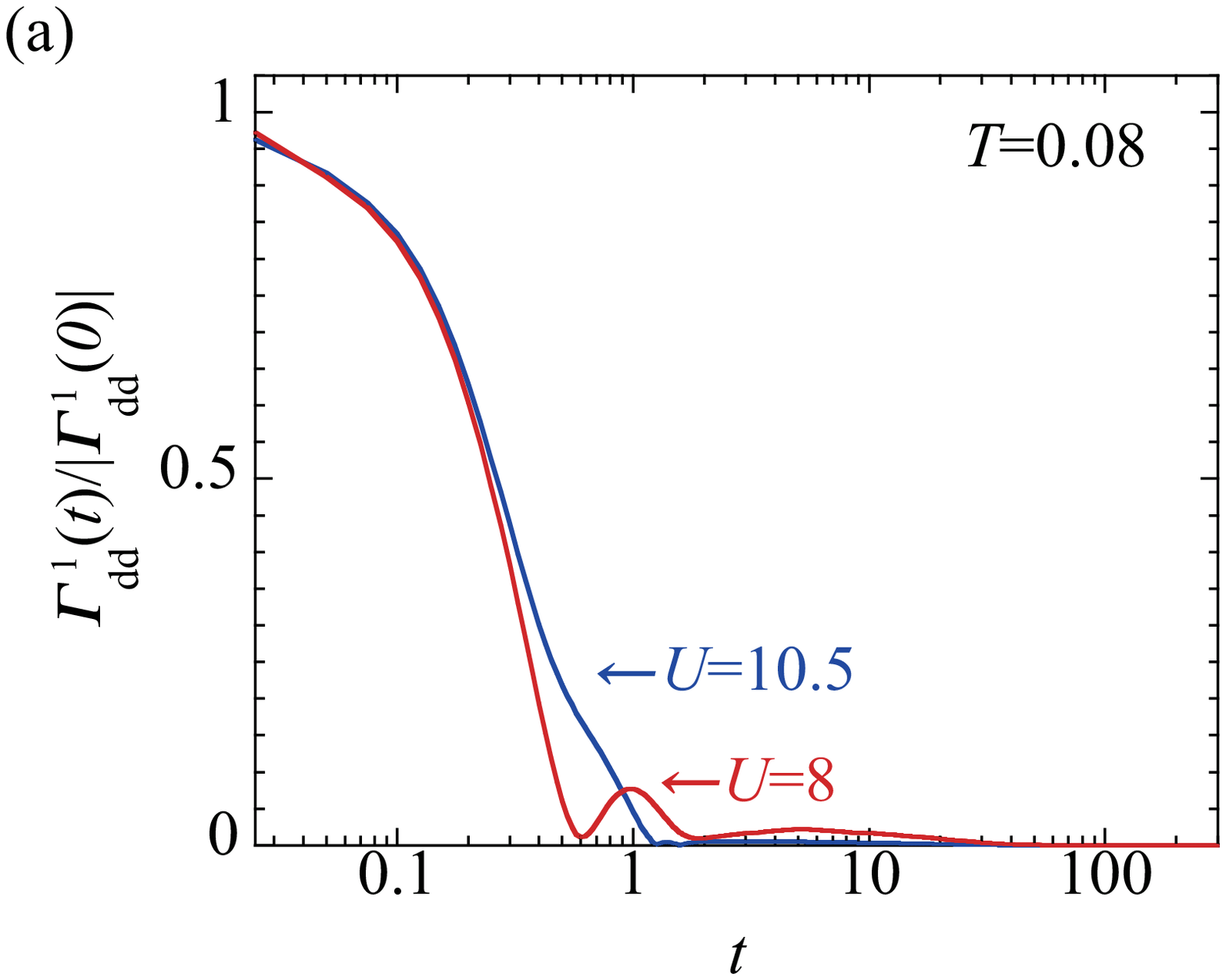}}
\vspace{-5.75cm}
\centerline{ \includegraphics[height=4.5in]{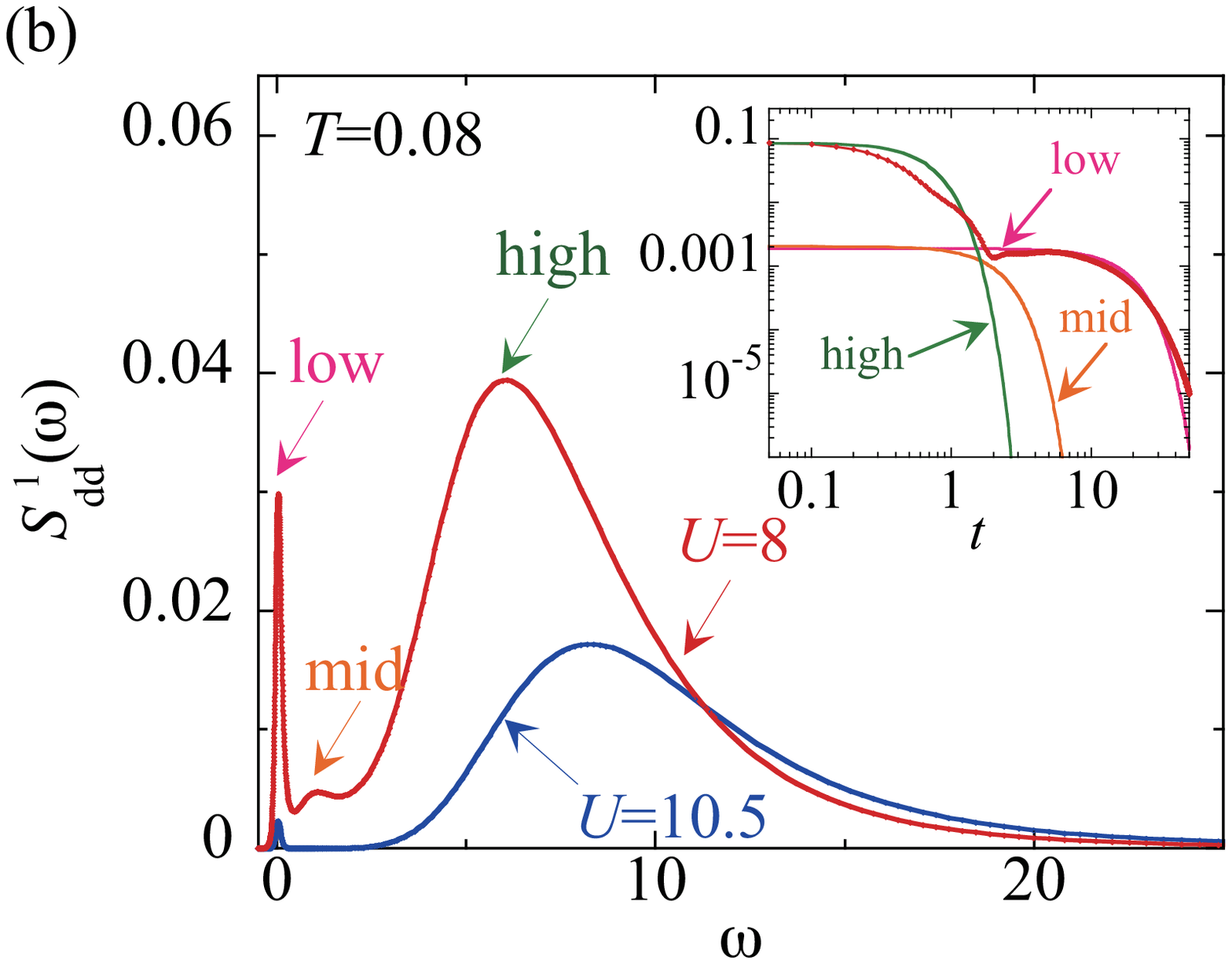}}
\vspace{-3.25cm}
\centerline{\includegraphics[height=4.5in]{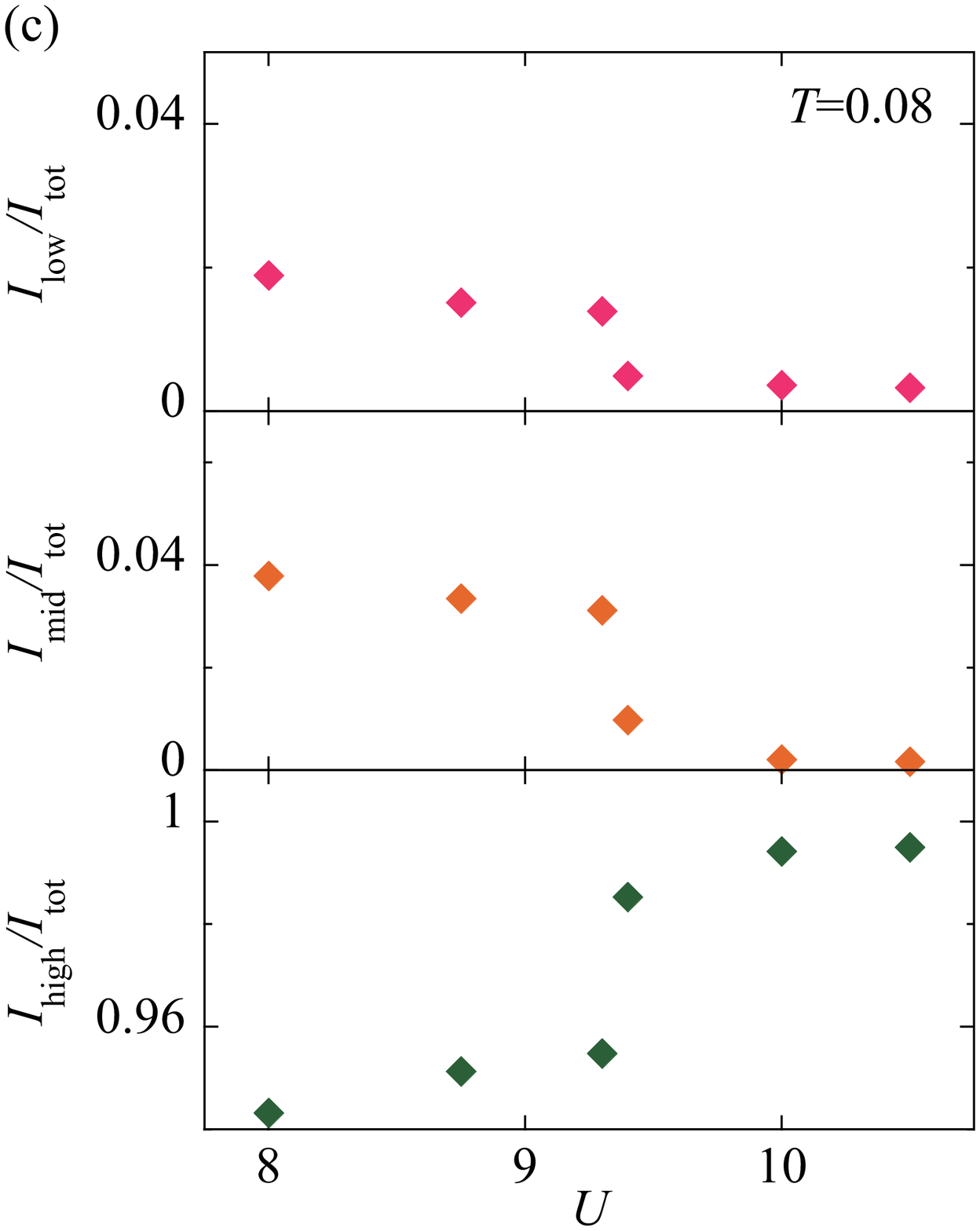} }
\vspace{-1.75cm}
\caption{(Color online) On-site dynamical correlation of doublon for two $U$'s (a) $\Gamma^{1}_{\rm oo'}(t)$.
 (b) $S^{1}_{\rm dd}(\omega)$
Inset: Diamonds show calculated data in $t$ and lines are the fitting results of three peaks around $\omega$=$0$ (``low"),
$\omega$$\sim$$1$ (``mid"), and $\omega$$\sim$$U$ (``high") at $U$=$8$. 
(c) $U$-dependence of the intensities of the three parts of $S^{1}_{\rm dd}(\omega)$ normalized by the total intensity.
}  
\label{fig:1}
\end{figure}

Next, we examine the variation of the on-site doublon dynamics.
As shown in Fig.~\ref{fig:1} (a), $\Gamma^{1}_{\rm dd}(t)$ decreases with time $t$ for both of $U$=$8$ and $10.5$.
However, this decrease is not monotonic, in particular at smaller $U$=$8$, which indicates larger fluctuation of doublon.
The relaxation time towards $\Gamma^{1}_{\rm dd}(t)$$=$$0$ is related with the doublon lifetime
and we expect a longer life time in the metallic phase than that in the insulating phase. 
The present result shows that the relaxation is slower in the long-$t$ part at $U$=$8$ than that at $U$=$10.5$ as expected.
The interesting characteristic is the two structures at $t$$\sim$$0.6$ and $2$, in particular in the metallic phase.
We investigate the complex $S^{1}_{\rm dd}(t)$, and find that the structure at $t$$\sim$$0.6$ corresponds to its position
that the phase is close to $\pi/2$ whereas that at $t $$\sim$$2$ corresponds the position with the phase $\pi$.

We also calculate the correlation in the frequency domain, ${S}^{1}_{\rm dd}(\omega)$, as shown in the main panel of Fig. \ref{fig:1} (b).
An important characteristic at $U$=$8$ is the three peaks in ${S}^{1}_{\rm dd}(\omega)$:  
one is the large broad peak around $\omega$=$U$ (``high"), another is the very sharp peak around $\omega$=$0$ (``low"),
and the last is the small peak at $\omega$$\sim$$1$ (``mid").
The broad peak around $\omega$=$U$ corresponds to the incoherent dynamics due to the excitations to the Hubbard band,
and this persists from the insulating to metallic phase.
The sharp peak around $\omega$=$0$ is due to the dynamics of coherent quasiparticles.
This one and the small peak at $\omega$$\sim$$1$ disappear in the insulating phase, as shown in the result at $U$=$10.5$.
We estimate the intensity of the three peaks $I_{\rm low,mid,high}$ and examine their changes with $U$.
The high-$\omega$ broad and the $\omega$$\sim$$0$ sharp peaks are fitted by Gaussian functions to define $S^{1,\rm high(low)}_{\rm dd}(\omega)$
and the remaining part is denoted as $S^{1,\rm mid}_{\rm dd}(\omega)$.
$I_{\rm low,mid,high}$ are the intensities of these three parts.
We calculate their ratio to the total intensity $I_{\rm tot}$, and the results are shown in Fig.~\ref{fig:1}~(c).
In the metallic phase, $I_{\rm low}/I_{\rm tot}$ and $I_{\rm mid}/I_{\rm tot}$ decrease with increasing $U$,
while $I_{\rm high}/I_{\rm tot}$ increases.
At the Mott transition point,  all the three intensities show a jump.
For larger $U$ in the insulating phase, only $I_{\rm high}/I_{\rm tot}$ takes a noticeable value.
We also calculate the real-time correlations $S^{1,\rm low,mid,high}_{\rm dd}(t)$ by the Fourier transformation 
and the results are shown in the inset of Fig.~\ref{fig:1} (b). 
One can see that the high-$\omega$ contribution is dominant at short $t$.
On the other hand, the long-$t$ dynamics is predonimated by only the low-$\omega$ part.
We find that the change from the short-$t$ to the long-$t$ behavior occurs around $t$$\sim$$2$.

In this letter, we have studied the dynamics of doublons in the triangular-lattice Hubbard model at half filling
by using the cellular dynamical mean field theory.  
We have demonstrated that a nearest-neighbor doublon-holon pair exhibits a strong attraction in the insulating phase.
The nearest-neighbor pairs of doublons, of holons, and of doublon and single occupied site show a repulsive correlation,
but they do not show as large enhancement as doublon-holon attraction.
Calculating the on-site doublon dynamics, we have found quantitatively that doublons in the metallic phase have a longer life time than that in the insulating phase. 
Our results of dynamical correlations show clear differences between in the metallic and in the insulating phases
and demonstrate complex dynamics of doublons in the metallic phase, which is associated with the several excitation process.

The authors are grateful to Kazumasa Hattori for helpful discussions. 
The present work is supported by MEXT Grant-in-Aid for Scientific Research No.25400359,
and by Next Generation Supercomputing Project, Nanoscience Program, MEXT, Japan.
Numerical computation was performed with facilities at Supercomputer Center at ISSP and Information Technology Center,
University of Tokyo.


\begin{thebibliography}{}
\bibitem{MT-exp-1}D. B. McWhan, et al., Phys. Rev. Lett. {\bf 27}, 941 (1971).
\bibitem{MTC-dcc-DMFT-1}G. Kotliar, E. Lange, and M. J. Rozenberg, Phys. Rev. Lett. {\bf 84}, 5180 (2000).
\bibitem{MT-critexp-1}P. Limelette, A. Georges, D. J$\acute{e}$rome, P. Wzietek, P. Metcalf, and J. M. Honig, Science {\bf 302}, 89 (2003).
\bibitem{MT-critexp-2}F. Kagawa, K. Miyagawa, and K. Kanoda, Nature (London) {\bf 436}, 534 (2005).
\bibitem{MI-crit-1}M. Imada, Phys. Rev. B {\bf 72}, 075113 (2005).
\bibitem{docc-scaling} P. S$\acute{\rm e}$mon and A.-M. S. Tremblay, Phys. Rev. B {\bf 85}, 201101(R) (2012).
\bibitem{docc-scaling-2} T. Sato, K. Hattori, and H. Tsunetsugu, Phys. Rev. B {\bf 86}, 235137 (2012).
\bibitem{rf-1} H. Kusunose, J. Phys. Soc. Jpn. {\bf 75}, 054713 (2006).
\bibitem{rf-2} T. Ohashi, N. Kawakami, and H. Tsunetsugu, Phys. Rev. Lett. {\bf 97}, 066401 (2006).
\bibitem{rf-7} T. Ohashi, T. Momoi, H. Tsunetsugu, and N. Kawakami,  Prog. Theor. Phys. Suppl. {\bf 176}, 97 (2008).
\bibitem{rf-3} B. Kyung, Phys. Rev. B {\bf 75}, 033102 (2007).
\bibitem{rf-4} M. J. Rozenberg, G. Kotliar, H. Kajueter, G. A. Thomas, D. H. Rapkine, J. M. Honig, and P. Metcalf, Phys. Rev. Lett. {\bf 75}, 105 (1995).
\bibitem{rf-5} J. Merino and R. H. McKenzie, Phys. Rev. B {\bf 61}, 7996 (2000).
\bibitem{rf-6} T. Sato, K. Hattori, and H. Tsunetsugu, J. Phys. Soc. Jpn. {\bf 81}, 083703 (2012). 
\bibitem{docc-op} C. Castellani, C. Di Castro, D. Feinberg, and J. Ranninger, Phys. Rev. Lett. {\bf 43}, 1957 (1979).
\bibitem{dc-3} H. Yokoyama, M. Ogata, and Y. Tanaka, J. Phys. Soc. Jpn. {\bf 75}, 114706 (2006).
\bibitem{dc-4} T. Watanabe, H. Yokoyama, Y. Tanaka, and J. Inoue, J. Phys. Soc. Jpn. {\bf 75}, 074707 (2006).
\bibitem{DMFT} A. Georges, G. Kotliar, W. Krauth, and M. J. Rozenberg, Rev. Mod. Phys. {\bf 68}, 13 (1996).
\bibitem{CDMFT} G. Kotliar, S. Y. Savrasov, G. P$\acute{\rm a}$lsson, and G. Biroli, Phys. Rev. Lett. {\bf 87}, 186401 (2001).
%
\bibitem{magphase-trian}P. Sahebsara and D. S$\acute{e}$n$\acute{e}$chal, Phys. Rev. Lett. {\bf 100}, 136402 (2008).
%
\bibitem{CTQMC} P. Werner, A. Comanac, L. de' Medici, M. Troyer, and A. J. Millis, Phys. Rev. Lett. {\bf 97}, 076405 (2006).
%
\bibitem{MEM} M. Jarrell and J.E. Gubernatis, Phys. Rep. {\bf 269}, 133 (1996).
%
\bibitem{phase}A. Liebsch, H. Ishida, and J. Merino, Phys. Rev. B {\bf 79}, 195108 (2009).

\end{thebibliography}
\end{document}